\documentclass{iopart}
\usepackage{latexsym,iopams}
\usepackage[dvips]{graphicx,color}
\usepackage{colortbl}
\usepackage{times}

\begin{document}
\bibliographystyle{unsrt}
\title[Extended 2d dilaton theories]{Extended 2d generalized dilaton gravity theories}

\author{R. O. de Mello}

\address{Institute for Theoretical Physics - IFT, S\~ao Paulo State University - UNESP, \\
Rua Pamplona 145, 01405-900, S.Paulo, SP, Brazil}
\ead{romello@ift.unesp.br}

\begin{abstract}
We show that an anomaly-free description of matter in (1+1) dimensions requires a deformation
of the 2d relativity principle, which introduces a non-trivial center in the 2d Poincar\'e
algebra. Then we work out the reduced phase-space of the anomaly-free 2d relativistic particle,
in order to show that it lives in a noncommutative 2d Minkowski
space.  Moreover, we build a Gaussian wave packet to show that a Planck length is well-defined
in two dimensions. In order to provide a gravitational interpretation for this
noncommutativity, we propose to extend the usual 2d generalized dilaton gravity models
by a specific Maxwell component, which gauges the extra symmetry associated with the center of the 2d
Poincar\'e algebra. In addition, we show that this extension is a high energy 
correction to the unextended dilaton theories that can affect the topology of space--time.
Further, we couple a test particle to the general extended dilaton models with the purpose of showing that
they predict a noncommutativity in curved space-time, which is locally described by
a Moyal star product in the low energy limit. We also conjecture a probable generalization of
this result, which provides a strong evidence that the noncommutativity is described by a certain star
product which is not of the Moyal type at high energies. Finally, we prove that the extended
dilaton theories can be formulated as
Poisson--Sigma models based on a nonlinear deformation of the extended Poincar\'e algebra.

\end{abstract}
\submitto{\CQG}

\section{Introduction}
\label{intro}
Since the discovery of the accelerated and flat universe, dilaton theories became strong
candidates for describing the effective theory of gravity in four dimensions. On the other hand, superstring theory is still afflicted with several theoretical problems and it has a lot of
difficulty to predict a quantum gravity phenomenology that can be tested by experiments. Nowadays, 2d dilaton gravity
theories are widely regarded as invaluable tools for researching
several aspects of quantum gravity. Indeed, they provide a unitary quantum theory of
gravity in a background independent formulation~\cite{kummer,kummer3,grumiller8}, with
a lot of application in black hole physics~\cite{hyun,youm2,cadoni}.

In this paper, we will show that the 2d special relativity principle must be formulated in
terms of the centrally extended 2d Poincar\'e group, in order to provide a consistent
description of a new class of 2d relativistic elementary systems, which the 2d
particle theory must allow for due to the non-trivial second cohomology of the 2d Poincar\'e
group. In Ref.~\cite{artigo}, we have already studied the extended Poincar\'e group in detail.
In particular, we have worked out all the unitary irreducible representations
(irrep's) of this group by the orbit method and we have provided a physical interpretation for a particular class of irrep's --- which corresponds to the new class of elementary systems that
we will consider here ---,
in terms of the anomaly-free relativistic particle. In this work, we will proceed working out the reduced phase-space of the anomaly-free relativistic
particle to show that it lives in a noncommutative 2d Minkowski space, in the absence of gravity. Further, we will build a Gaussian wave packet to probe this noncommutativity
and to show that a Planck length is well-defined in two dimensions.  

The main result of the present work is the introduction of a new class of theories, which
extends the usual 2d generalized dilaton gravity models, in order to provide a gravitational
interpretation for this noncommutative structure. We will propose a specific class of
dilaton--Maxwell theories, which includes an additional U(1) sector that gauges the extra
symmetry introduced by the central extension of the Poincar\'e group. It will
be shown that this U(1) gauge sector can be physically interpreted as a high energy correction to the unextended dilaton theories that can affect the topology of space--time. Moreover, we will prove that this new class of models can also be formulated
as a family of nonlinear gauge theories~\cite{ikeda}, in terms of Poisson--Sigma
models~\cite{schaller2} based
on a deformation of the centrally extended Poincar\'e algebra.

This paper is organized as follows. From Sec.~\ref{sec:1} to Sec.~\ref{sec:3} we will present
the physical motivation for introducing the extended dilaton models, briefly
reviewing some relevant results of Ref.~\cite{artigo}, in order to make
the exposition as self-contained as possible. In Sec.~\ref{sec:1}, we will explain what kind of anomaly
arises in the quantization of
most 2d relativistic elementary systems, unless we formulate 
the 2d special relativity principle in terms of the centrally extended Poincar\'e group and, in
Sec.~\ref{sec:2}, we will illustrate the formalism with the quantum
anomaly-free 2d relativistic particle. In Sec.~\ref{sec:3}, we will 
build a Gaussian wave packet to
probe the noncommutativity which is produced in 2d Minkowski space by this deformation
of the 2d Poincar\'e symmetry.
In Sec.~\ref{sec:4}, we will introduce the
extended 2d generalized dilaton gravity theories and show how they predict a small noncommutativity
in curved space-time in the low energy limit. Then
we will show that the extended models can be formulated as
Poisson--Sigma models. In Tab.~\ref{tab:selec} we will provide an updated list of the main
extended models. Finally, in Sec.~\ref{sec:5} we will discuss our results,
besides pointing out other future developments.

\section{Anomalous elementary systems and the special relativity principle in 1+1 dimensions}
\label{sec:1}
In this section, we will show that the quantization of most relativistic elementary systems in
1+1 dimensions looks anomalous due to the non-triviality of the second cohomology group of the
two-dimensional Poincar\'e group $\mathcal{P}$. Further, an anomaly-free formulation of the
2d special relativity principle is possible only if we deform the symmetry group into the
centrally extended Poincar\'e group $\bar{\mathcal{P}}$. As a consequence of this deformation, all matter fields become always subjected to a characteristic
interaction, even in the absence of gravity and any other Yang-Mills potentials.

Let us start recalling that the special relativity principle requires that the laws of physics transform covariantly under
the Poincar\'e group. This is usually a quite straightforward statement, except in 1+1
dimensions, where we have to be carefull about a mathematical subtlety that has been shown by
Bargmann: the fact that, exclusively in the two-dimensional case, the second cohomology group
of the Poincar\'e group is non-trivial. Indeed, it turns out that the second cohomology group
of the 2d Poincar\'e group is one-dimensional, $H^{2}_{0}(\mathcal{P},\Re)=\Re$, and
parametrized by the central charge $B\in\Re$.

It is well known that the theory of cohomology of Lie groups is closely related to the theory of
central extensions of Lie groups. In particular, it can be shown that a non-trivial two
cocycle $\xi(g',g)$ in the second cohomology group of a Lie group $G$ (where $g',g\in G$)
corresponds to a left-invariant
two form $\omega^{(2)}$ in the second Chevalley-Eilenberg cohomology group $E^{2}(G)$, which defines a
non-trivial central extension of the Lie algebra of $G$ by $\Re$, denoted by
$\bar{\mathfrak{g}}$ (for further details concerning the theory of cohomology of Lie groups
and Lie algebras, see Ref.~\cite{azcarraga}).

In the case of the 2d Poincar\'e algebra, it is clear that it can be
non-trivially extended in precisely one way, expressed by the extended Poincar\'e algebra $\bar{\textrm{\i}}^{1}_{2}$ given by
\begin{equation}\label{eq:algpoincest}
\fl \lbrack P_{a},J\rbrack = \sqrt{-h}\epsilon^{\verb+ +b}_{a}P_{b}\textrm{,}\qquad
\lbrack P_{a},P_{b}\rbrack=B\epsilon_{ab} I\textrm{,}\qquad\textrm{and}\qquad
\lbrack P_{a},I\rbrack  =  \lbrack J,I\rbrack=0\textrm{,}
\end{equation}
where $P_{a}$ and $J$ are respectively the generators of translations and Lorentz
transformations, while $I$ is the central generator and $B$ is the central charge. The $\epsilon$-tensor is such that $\epsilon_{ab}$ is dimensionless, $\epsilon^{01}=-\epsilon_{01}=1$, and the indices $(a,b)$ are raised and lowered by the metric
$h_{ab}=\mathrm{diag}(1,-1)$, with $h:=\mathrm{det}\,h_{ab}=-1$. 

Throughout this paper we shall
stick to the same conventions of Ref.~\cite{artigo}, in order to be consistent with the group
theoretic framework developed therein. In particular, we shall adopt units where only the speed
of light is set to $c=1$ (the Planck constant is not set to unit) and such that the 2d
Minkowski space coordinates have dimension $[q^{a}]=L$.
Moreover, the algebra $\bar{\textrm{\i}}^{1}_{2}$ is regarded to be anti-homomorphic to the
Lie algebra of the generators of the left action of the Poincar\'e group
on Minkowski space $q'^{b} = \theta^{b}+\Lambda(\alpha)^{b}_{\verb+ +a}q^{a}$,
where Lorentz transformations are given by $\Lambda(\alpha)^{a}_{\verb+ +b}=\delta^{a}_{\verb+ +b}\cosh\alpha+\sqrt{-h}\,\epsilon^{a}_{\verb+ +b}\sinh\alpha$. It follows that the boost parameter $\alpha$ is dimensionless
and the translation parameters have dimension $[\theta^{a}]=L$, so that the generator
$J$ is dimensionless and $[P_{a}]=L^{-1}$.

It will be shown below that the central charge $B$ has
dimension $[B]=L^{-2}\times[\hbar]$, so that Eq.~(\ref{eq:algpoincest}) implies that the
dimension of the central generator is inverse of action $[I]=[\hbar]^{-1}$.
On the other hand, in the group
theoretic framework developed in Ref.~\cite{artigo} the metric $h_{ab}$ is defined by the translation sector of the invariant Casimir operator $h^{ab}P_{a}P_{a}-2(B/\sqrt{-h})J I$, so that it has the unusual dimension $[h_{ab}]=L^{-2}$. As a result, we will often employ the factor $\sqrt{-h}=1$ (which carries one unit of dimension $L^{-2}$), in order to get the correct dimensions in several equations whenever we
raise and lower indices by the metric $h_{ab}$.

The group law of the extended Poincar\'e group  $g''(\theta''^{a},\alpha'',\beta'')\    =\    g'(\theta'^{a},\alpha',\beta')\   g(\theta^{a}\ ,\\\alpha,\beta)$
is determined by the group parametrization
$g(\theta^{a},\alpha,\beta)=\exp(\theta^{a}P_{a})\exp(\alpha J)\exp(\beta I)$ and Eq.~(\ref{eq:algpoincest}); it is given by
$\theta''^{b} = \theta'^{b}+\Lambda(\alpha')^{b}_{\verb+ +a}\theta^{a}$, $\alpha'' =
\alpha'+\alpha$, and $\beta'' = \beta'+\beta+(B/2)\theta'^{c}\epsilon_{cb}\Lambda(\alpha')^{b}_{\verb+ +a}\theta^{a}$, where $\beta$ is the group parameter corresponding to the central
generator $I$, so that it has dimension $[\beta]=[\hbar]$. Applying the standard theory of
central extensions of Lie groups~\cite{azcarraga}, we can read the non-trivial two cocycle
$\xi_{B}(g',g)$ in the second cohomology group $H^{2}_{0}(\mathcal{P},\Re)$ from the composition law for
$\beta$ above, which is given by
$\xi_{B}(g',g)=(B/2)\theta'^{c}\epsilon_{cb}\Lambda(\alpha')^{b}_{\verb+ +a}\theta^{a}$
for $g(\theta^{a},\alpha)\in \mathcal{P}$. This non-trivial two cocycle corresponds to the
left-invariant two form $\omega^{(2)}_{(B)}=(B/2)\epsilon_{ab}\rmd\theta^{a}\wedge\rmd\theta^{b}$ in the second Chevalley-Eilenberg cohomology group $E^{2}(\mathcal{P})$, which determines the non-trivial central extension given by Eq.~(\ref{eq:algpoincest}).

As a particular consequence of the special relativity principle, the elementary particles --- 
i.e. all types of matter fields (scalar,
tensor, and spinorial) --- should belong to the irrep's of the Poincar\'e group, being regarded as
relativistic elementary systems in this sense. On the other
hand, in quantum mechanics they should also correspond to pure states, which are represented by
rays, so that the Poincar\'e symmetry operators should be realized by unitary ray operators.
So, in order to realize the 2d special relativity
principle at the quantum level, we must consider the projective
unitary ray representations of the 2d Poincar\'e group $\mathcal{P}$,
$U(g')U(g)=\omega_{B}(g',g)U(g'g)$ for $g', g\in\mathcal{P}$, which are defined by the local factors 
$\omega_{B}(g',g)=\exp[(i/\hbar)\xi_{B}(g',g)]$. In particular, it follows from this
expression for $\omega_{B}(g',g)$ that the dimension of the central charge is fixed to
$[B]=L^{-2}\times[\hbar]$.

It is easy to see that the inequivalent classes of unitary ray representations of
$\mathcal{P}$ are associated with the non-equivalent two cocycles $\xi_{B}(g',g)$ in $H^{2}_{0}(\mathcal{P},\Re)$.
It follows that the Hilbert space of all possible states of a 2d elementary relativistic
system --- of some given type (spinorial, etc.) --- is decomposed
into coherent subspaces characterized by the central charge $B$. Since states corresponding to two different values of $B$ cannot be coherently mixed, there are no states with a central
charge spectrum and therefore no transitions can occur among distinct subspaces. This result
is analogous to the well known Bargmann mass superselecion rule, which applies in the case of the Galilei group in non-relativistic quantum mechanics~\cite{azcarraga}.

Thus the unitary ray representations of $\mathcal{P}$ actually define a multiplicity of inequivalent relativistic elementary systems, well separated by a superselection rule. This means that
the symmetry principle alone is unable to determine any particular coherent subspace of the
total Hilbert space
for describing a given elementary system, since it doesn't fix any value for $B$.
The introduction of non-trivial elementary systems (corresponding to $B\ne 0$) is an unusual
feature, since in four dimensions all ray representations of the Poincar\'e group are equivalent to the trivial one.
Already this ambiguity in the definition of an elementary system --- which is not found in
4d --- seems to indicate that the 2d Poincar\'e group $\mathcal{P}$ is not the most
physically suitable symmetry group in two dimensions.

In order to provide a physical interpretation for all these elementary systems, we will
now consider their classical limit. At the classical level, the irreducibility condition
on the representations of the symmetry group is naturally translated into a transitivity one, so that the classical relativistic elementary systems
should correspond to homogeneous symplectic manifolds (HSM's) with respect to the
Poincar\'e group (for
further details concerning the classification of
the relativistic 2d elementary systems, see Ref.~\cite{artigo} and the references therein). On the other hand, in the
Lagrangean formalism the special relativity principle should be implemented as an invariance of the action
under a rigid Poincar\'e transformation. However, in two dimensions there is a connection
between the second cohomology of the Poincar\'e group and the quasi-invariance of the action that must be properly accounted for.

In fact, due to the L\'evy--Leblond theorem~\cite{azcarraga} all the inequivalent Lagrangians
$L_{B}$ which are quasi-invariant (i.e. invariant up to a total derivative) under the transformations of $\mathcal{P}$ are classified by the second cohomology
of the 2d Poincar\'e group. This means that each non-trivial two cocycle $\xi_{B}(g',g)$ in $H^{2}_{0}(\mathcal{P},\Re)$ must generate a quasi-invariant term in $L_{B}$. Let us see how it works
in the case of a relativistic point particle, which is usually described by the Lagrangean $L_{0}=-m(-h)^{-1/4}\sqrt{\dot{q}^{2}}$. The Lagrangean $L_{0}$ is invariant under $\mathcal{P}$, so
$\xi_{B}(g',g)$ must correspond to an extra term in $L_{B}=L_{0}+L_{WZ}$, which will be
denoted by $L_{WZ}$.
In order to work out the form of $L_{WZ}$, we will consider the action of $\mathcal{P}$ on the
homogeneous space $H=\mathcal{P}/SO(1,1)$, which can be identified to the 2d Minkowski space
by doing $q^{a}=\theta^{a}$.    

Thus, the left-invariant two form defining the central extension
$\bar{\textrm{\i}}^{1}_{2}$ of Eq.~(\ref{eq:algpoincest}) is well-defined on Minkowski space
$H$ and it is given by $\omega^{(2)}_{(B)}=(B/2)\epsilon_{ab}\rmd q^{a}\wedge\rmd q^{b}$. This two
form must be exact, since it is closed and $H$ has a trivial de Rham cohomology, so we may
write $\omega^{(2)}_{(B)}=d\beta^{(1)}_{(B)}$ where $\beta^{(1)}_{(B)}=(B/2)\varepsilon_{ab}q^{a}dq^{b}$. On the other hand, $\omega^{(2)}_{(B)}$ is a non-trivial Chevalley--Eilenberg (CE) two
cocycle, hence the potential one form $\beta^{(1)}_{(B)}$ cannot be left-invariant. Indeed, it
is easy to see that $\beta^{(1)}_{(B)}$ is quasi-invariant under a finite Poincar\'e transformation
$g=g(\theta^{a},\alpha)$ acting on the left, for $\delta \beta^{(1)}_{(B)}=d\Delta_{(B)}(q;g)$
where $\Delta_{(B)}(q;g)=(B/2)\theta^{a}\varepsilon_{ab}\Lambda(\alpha)^{b}_{\verb+ +c}q^{c}$,
consistently with the Lie derivative $L_{\theta^{a}T^{R}_{a}+
\alpha T^{R}_{2}}\beta^{(1)}_{(B)}=
d((B/2)\theta^{a}\varepsilon_{ab}q^{b})$ where $T^{R}_{A}\equiv (P_{a},J)$ with
$A\in \{0,1,2\}$ denote
the generators of the left action of $\mathcal{P}$ on $H$ (which are right-invariant vector
fields).

Actually, the Lie derivatives $L_{T^{R}_{A}}\beta^{(1)}_{(B)}$ must be
exact due to the left-invariance of $\omega^{(2)}_{(B)}$, since $L_{T^{R}_{A}}\omega^{(2)}_{(B)}=dL_{T^{R}_{A}}\beta^{(1)}_{(B)}=0$. In fact, the calculation above shows that $L_{T^{R}_{A}}\beta^{(1)}_{(B)}=d\Delta_{A}$ with $\Delta_{a}=(B/2)\varepsilon_{ab}q^{b}$ and $\Delta_{2}=0$, so that $\beta^{(1)}_{(B)}$ associates a function $\Delta_{A}$
on $H$ to each $T^{R}_{A}$. Moreover, the identity $[L_{X},L_{Y}]=L_{[X,Y]}$ implies
\begin{equation}\label{eq:wesszumino}
d(L_{T^{R}_{A}}\Delta_{B}-L_{T^{R}_{B}}\Delta_{A}-\Delta_{[A,B]})=0,
\end{equation}
which will be called
Wess--Zumino (WZ) consistency condition in analogy to a similar structure that holds in the
context of anomalous gauge theories. Thus $\beta^{(1)}_{(B)}$ is a WZ one form by definition,
since it is the potential one form of a non-trivial CE two cocycle.   

The WZ one form is well-defined on the 1-jet bundle $\Re\times T(H)$ with
local coordinates $(\tau,q^{a},\dot{q}^{a})$. So, we can get the extra term in the Lagrangean
$L_{B}$ by pulling $\beta^{(1)}_{(B)}$ back to the world line of the particle by the section
$s : \tau\to s(\tau)=(q^{a}(\tau),\dot{q}^{a}(\tau)=dq^{a}(\tau)/d\tau)$. As a result we
get $L_{WZ}:=s^{*}(\beta^{(1)}_{(B)})=-(B/2)\varepsilon_{ab}\dot{q}^{a}q^{b}$, which will be called Wess--Zumino term. Hence, the
Lagrangean is quasi-invariant under a 2d Poincar\'e transformation $L_{B}(q'^{a},\dot{q}'^{a})-
L_{B}(q^{a},\dot{q}^{a})=d(\Delta_{(B)}(q;g))/d\tau$, so that $\xi_{B}(g',g)=\Delta_{(B)}(q';g')-\Delta_{(B)}(q;g'g)+\Delta_{(B)}(q;g)$, in perfect agreement with the L\'evy--Leblond theorem.

Nevertheless, calculating the conserved Noether charges $\mathcal{N}_{2}=p_{a}\varepsilon^{a}_{\verb+ +b}q^{b}$ and $\mathcal{N}_{a}=p_{a}-\Delta_{a}$, we note that they close a Poisson
bracket realization of the extended algebra $\bar{\textrm{\i}}^{1}_{2}$ of Eq.~(\ref{eq:algpoincest}), given by $\{\mathcal{N}_{a},\mathcal{N}_{2}\}=\varepsilon_{a}^{\verb+ +b}\mathcal{N}_{b}$ and $\{\mathcal{N}_{a},\mathcal{N}_{b}\}=B\varepsilon_{ab}$, rather than the 2d Poincar\'e
algebra. As a consequence, when we apply a quantization rule to this algebra of charges, in order to
impose the Poincar\'e symmetry at the quantum level, we will get an anomaly for the non-trivial systems ($B\ne 0$). The anomaly can be traced back to
the $\Delta_{a}$ terms in the charges $\mathcal{N}_{a}$, which were introduced by the WZ one
forms $\beta^{(1)}_{(B)}$ stemming from the
non-trivial two cocycles $\xi_{B}(g',g)$ in $H^{2}_{0}(\mathcal{P},\Re)$.

On the one hand, it is clear that the central charge $B$ is a free parameter that must be fixed at the outset,
due to the existence of a non-trivial two cocycle in $H^{2}_{0}(\mathcal{P},\Re)$; it is
characteristic of the 2d relativistic elementary systems, just like the mass $m$. On the other
hand, although we do not expect any anomaly to appear in the trivial case, the non-trivial systems constitute physical possibilities that cannot be ruled out, since they do not
violate the 2d Poincar\'e symmetry. Therefore, instead of assuming $B=0$, it seems much wiser
to search for an anomaly-free formulation of the 2d particle theory, which applies for all
values of $B$.

Obviously, the idea is to take the algebra of charges for the fundamental symmetry. Since
$H^{2}_{0}(\bar{\textrm{\i}}^{1}_{2},\Re)=0$ (see Ref.~\cite{artigo} for a proof), if we
substitute a left-invariant one form $\bar{\beta}^{(1)}_{(B)}$ for the WZ one form, we will
not get an anomaly anymore. This can be done by considering the action of the extended
Poincar\'e group $\bar{\mathcal{P}}$ on the homogeneous space
$\bar{H}=\bar{\mathcal{P}}/SO(1,1)$ with coordinates identified to
$(q^{a},\chi)=(\theta^{a},\beta)$, where $\chi$ is an extra coordinate transforming
as $\chi'=\chi+\beta+(B/2)\theta^{a}\epsilon_{ab}\Lambda^{b}_{\verb+ +c}q^{c}$. Then, we can
naturally take for $\bar{\beta}^{(1)}_{(B)}$ the left-invariant Maurer-Cartan one form associated with the group parameter
$\beta$, with a global sign added to it. The one form $\bar{\beta}^{(1)}_{(B)}$ is well-defined on $\bar{H}$
and it is given by $\bar{\beta}^{(1)}_{(B)}=(B/2)\varepsilon_{ab}q^{a}dq^{b}-d\chi$. Of course, now
we have $L_{\theta^{a}\bar{T}^{R}_{a}+
\alpha\bar{T}^{R}_{2}+\beta\bar{T}^{R}_{3}}\bar{\beta}^{(1)}_{(B)}=0$, where $\bar{T}^{R}_{A}\equiv (P_{a},J,I)$ with
$A\in \{0,1,2,3\}$ denote
the generators of the left action of $\bar{\mathcal{P}}$ on $\bar{H}$, so that $\bar{\beta}^{(1)}_{(B)}$ is not a WZ one form.

We proceed considering the 1-jet bundle $\Re\times T(\bar{H})$ with
local coordinates $(\tau,q^{a},\chi,\dot{q}^{a},\dot{\chi})$, so that we get
the anomaly-free Lagrangean $\bar{L}_{B}=L_{0}+L_{WZ}-\dot{\chi}$ by adding to the original
Lagrangean $L_{0}$ the pull-back $\bar{s}^{*}(\bar{\beta}^{(1)}_{(B)})=-(B/2)\varepsilon_{ab}\dot{q}^{a}q^{b}-\dot{\chi}$ by the section
$\bar{s} : \tau\to \bar{s}(\tau)=(q^{a}(\tau),\chi(\tau),\dot{q}^{a}(\tau)=dq^{a}(\tau)/d\tau,\dot{\chi}(\tau)
=d\chi(\tau)/d\tau)$. The third term in $\bar{L}_{B}=L_{B}-\dot{\chi}$ introduces an auxiliary
degree of freedom $\chi$ with dimension of action and neutralizes the WZ term, since their variations compensate each other. The anomaly has been cancelled indeed, for now we have a
Lagrangean $\bar{L}_{B}$ which is invariant under $\bar{\mathcal{P}}$, while the four Noether
charges $\{\mathcal{N}_{a},\mathcal{N}_{2},\mathcal{N}_{3}=-1\}$ realize the extended
Poincar\'e algebra $\bar{\textrm{\i}}^{1}_{2}$.

Therefore, in order to describe the non-trivial elementary systems in an anomaly-free fashion
we must formulate the special relativity principle in terms of the extended Poincar\'e group
$\bar{\mathcal{P}}$. Note that the extended Poincar\'e group is not unique. Actually there is an
infinite number of inequivalent extended Poincar\'e groups $\bar{\mathcal{P}}_{(B)}$,
characterized by different values of the central charge $B$. Further, for a fixed $\bar{\mathcal{P}}_{(B)}$ the corrected special relativity principle formulated above can uniquely determine the Hilbert space of any
type of elementary system at the quantum level, thus solving the aforementioned ambiguity
problem. In fact, all ray representations of $\bar{\mathcal{P}}_{(B)}$ are
equivalent to the trivial one, since $H^{2}_{0}(\bar{\mathcal{P}}_{(B)},\Re)=0$. It is worth
mentioning that, although the transformation properties of the classical systems are related
to the extension $\bar{\mathcal{P}}$ of $\mathcal{P}$ by $\Re$, the quantum systems are
connected to the extension $\tilde{\mathcal{P}}$ of $\mathcal{P}$ by U(1). The groups $\bar{\mathcal{P}}$ and $\tilde{\mathcal{P}}$ are locally isomorphic and we will make no distinction among them in the following sections. 

To conclude, we remark that the two form $\omega^{(2)}_{(B)}$ is analogous to a constant electric force field, which is canonically defined throughout the 2d Minkowski space. Indeed, the WZ term $L_{WZ}$
has the structure of an interaction term, so that the WZ one form $\beta^{(1)}_{(B)}$
acts like an external applied U(1) potential, which drives all particles into an uniformly accelerated relativistic motion. Hence, $\beta^{(1)}_{(B)}$ will be called a WZ potential whenever we are
in a more physical context. This peculiar interaction has already received a gauge theoretic formulation in the cases of the Jackiw--Teitelboim model as well as of the conformally transformed string inspired
dilaton gravity~\cite{cangemi2,jackiw2,jackiw3}. However, in our approach this
interaction is caused by a WZ potential and, in Sec.~\ref{sec:4}, we will provide a contrasting
gravitational interpretation for it, which applies for general 2d dilaton gravity models.

\section{The anomaly-free 2d relativistic particle}
\label{sec:2}

In this section, we will introduce the quantum anomaly-free 2d relativistic
particle, as an example of the anomaly-free formulation for matter in the absence of gravity. Although the same general guidelines ---
which basically involve the coupling to the characteristic interaction that was explained in Sec.~\ref{sec:1} --- should
also apply to general matter fields, here we will only consider point particles. We will do this not just for simplicity but mainly because, in the next section, a point particle will turn out to be a very
suitable tool for
probing the noncommutativity which is produced in 2d Minkowski space by that peculiar
interaction.

In the trivial case ($B=0$), the description of the 2d relativistic particle is merely
standard and no noncommutativity shows up. However, for the non-trivial systems ($B\ne 0$)
in the absence of gravity, we are showing below that the anomaly cancellation, which
was explained in the previous section, induces a noncommutative structure in 2d Minkowski
space. Hence, unless otherwise stated, from now on we will assume that the central charge is
non-vanishing, since this is the most physically interesting situation. 

The Hamiltonian form of the Lagrangean $\bar{L}_{B}$, which describes the
anomaly-free 2d relativistic particle (cf. Sec.~\ref{sec:1}), is given by 
\begin{equation}\label{eq:acaoiltonia2}
S[q^{a},\chi,p_{b},\pi,u^{m},v^{n}]=\int_{\tau_{1}}^{\tau_{2}}\rmd\tau(p_{a}\dot{q}^{a}+\pi\dot{\chi}-u^{m}\phi_{m}-v^{n}C_{n})
\textrm{,}
\end{equation}
where $(p_{b},\pi)$ are the momenta canonically conjugate to the configuration space
coordinates $(q^{a},\chi)$, $(u^{m},v^{n})$ for $m,n\in\{1,2\}$ are Lagrange multipliers, and
$\tau$ parametrizes the world line of the particle. Calculating the Hessian of $\bar{L}_{B}$, we conclude that there
are two primary constraints: $\phi_{1}=\pi+1$ and $\phi_{2}=\tilde{p}^{a}\tilde{p}_{a}-m^{2}$, where $\tilde{p}_{a}=p_{a}-(B/2)\pi\epsilon_{ab}q^{b}$ is the kinematical momentum bivector. Further, since the canonical Hamiltonian is null and $\{\phi_{1},\phi_{2}\}=0$, we
deduce that there are no secondary constraints and that $\phi_{m}$ are
first class. The constraints $C_{n}$ are canonical gauge conditions to be discussed below.
In addition, calculating $\{\phi_{1},\chi\}=-1$, we discover that $\chi$ is not
an observable, so it is an unphysical gauge degree of freedom.

The reduced phase--space is obviously a two dimensional symplectic manifold, since there are
two first class constraints on the six dimensional phase--space with coordinates
$(q^{a},\chi,p_{b},\pi)$. In Ref.~\cite{artigo} we have already worked out the
reduced phase--space of this system by group theoretic methods. We have discovered that
it is parametrized only by the position coordinates $q^{a}$, independently of any gauge fixation, and that it is endowed with a noncommutative structure. It is worth mentioning that we have
also found that it corresponds to a coadjoint orbit of $\bar{\mathcal{P}}$ which is not a HSM
for $\mathcal{P}$ (cf. Subsec.~A of Sec.~III in Ref.~\cite{artigo}), which means that
the anomaly-free particle is a new kind of 2d relativistic elementary system.   
In the present work, we will work out the reduced phase--space by gauge fixation, which has a
clearer physical interpretation.

So let us fix the scale of $\tau$ by
imposing the unusual canonical gauge condition $C_{2}=\tilde{p}(\tau)-\tilde{p}(\tau_{0})-B(\tau-\tau_{0})$, where $\tilde{p}(\tau_{0})$ is a constant and $\tilde{p}(\tau):= \tilde{p}_{1}(\tau)$ is the kinematical momentum. This is a sensible choice, inasmuch as the
particle is accelerated by a constant physical force equal to $B$. In this formulation,
the usual condition $q^{0}=\tau$ appears (assuming $q^{0}(\tau_{0})=\tau_{0}$) as a Hamilton equation, together
with $q^{1}(\tau)=q^{1}(\tau_{0})+\Delta\mathcal{E}/B$, where $\Delta\mathcal{E}=\sqrt{m^{2}\!+\!\tilde{p}(\tau)^{2}}\! - \! \sqrt{m^{2}\!  +\!  \tilde{p}(\tau_{0})^{2}}$ is the variation
of the relativistic energy of the particle, which is given by $\mathcal{E}(\tau):=-\tilde{p}_{0}(\tau)$.   

On the other hand, the gauge degree of freedom $\chi$ can be naturally interpreted as the
phase of the particle's wave function, by imposing the canonical gauge condition
$C_{1}=\chi-S(q^{1},\tau)$, where $S(q^{1},\tau)=S(E(q^{1}),\tau)$ is the action
function of the theory, which can be expressed in terms of the
variable $E:=B q^{1}$ by
\begin{equation}\label{eq:acfunc}
S(E,\tau)=\frac{(\tilde{p}(\tau_{0})-2E+\Delta\mathcal{E})\Delta\mathcal{E}}{2B}+
\frac{m\Delta t'}{2}+\left(E-\frac{\mathcal{E}}{2}\right)\Delta\tau,
\end{equation}
with proper time interval $\Delta t'= (m/B)(\mathrm{arsinh}[\tilde{p}(\tau)/m]-\mathrm{arsinh}[\tilde{p}(\tau_{0})/m])$ and $\Delta\tau = \tau-\tau_{0}$. It is straightforward to check that
the Hamilton equations imply that
the Eq.~(\ref{eq:acfunc}) is a solution for the relativistic Hamilton-Jacobi equation
$H(\partial S/\partial E,E,\tau)-\partial S/\partial\tau\approx 0$ associated
with the Hamiltonian $H(q,E,\tau)=E-(\tilde{p}/\mathcal{E})(Bq+E)$, where $q:=q^{0}-q^{1}$,
which will be discussed below.

We can get the Hamilton equations by performing the variations of the
action~(\ref{eq:acaoiltonia2}) with respect to $q^{a}$, $\chi$, $p_{b}$, $\pi$, $u^{m}$,
and $v^{n}$. The result is
\begin{eqnarray}\label{eq:hameqs}
-\dot{p}_{0}-u^{2}\left[B\pi p_{1}-\frac{(B\pi)^{2}}{2}q_{0}\right]+
\frac{B\pi}{2}v^{2}  =  0,& &\nonumber\\
-\dot{p}_{1}-u^{2}\left[B\pi p_{0}+\frac{(B\pi)^{2}}{2}q^{1}\right]+v^{1}\frac{\partial S}{\partial q^{1}} = 0,& &\nonumber\\
\dot{q}^{0}-u^{2}(2p_{0}+B\pi q^{1})  = 0,& &\nonumber\\
\dot{q}^{1}-u^{2}(-2p_{1}+B\pi q^{0})-v^{2} = 0,& &\nonumber\\ -\dot{\pi}-v^{1} = 0,& &\nonumber\\
\dot{\chi}-u^{1}-u^{2}\left(-Bp_{a}\varepsilon^{a}_{\verb+ +b}q^{b}-\frac{B\pi}{2}q^{a}q_{a}
\right)+\frac{B}{2}v^{2}q^{0} = 0,& &\nonumber\\
\pi+1  \approx  0,& &\nonumber\\
\tilde{p}  \approx  \tilde{p}(\tau_{0})+B(\tau-\tau_{0}),& &\nonumber\\
p_{a}p^{a}-B\pi p_{a}\varepsilon^{a}_{\verb+ +b}q^{b}-
\frac{(B\pi)^{2}}{4}q^{a}q_{a}-m^{2} \approx  0,\textrm{ and}& &\nonumber\\
\chi-S(q^{1},\tau)  \approx  0.& &
\end{eqnarray}
Note that $\{C_{1},\phi_{1}\}=1$, $\{C_{2},\phi_{2}\}\approx -2B\mathcal{E}\not\approx 0$, and
$\{C_{2},\phi_{1}\}=0$, so that
$\textrm{det}\{\phi_{m},C_{n}\}\approx -2B\mathcal{E}\not\approx 0$, what shows that the
canonical gauge conditions are effective.

The consistency conditions on the constraints $\phi_{m}$, given by 
$\partial\phi_{m}/\partial\tau+\{\phi_{m},H_{C}\}+u^{n}\{\phi_{m},\phi_{n}\}+
v^{n}\{\phi_{m},C_{n}\}\approx 0$, where $H_{C}=0$ is the vanishing canonical Hamiltonian associated
with the Lagrangean $\bar{L}_{B}$ and $m,n\in\{1,2\}$, yield $v^{m}\approx 0$. The consistency
condition on $C_{2}$, given by $\partial C_{2}/\partial \tau+\{C_{2},H_{C}\}+u^{m}\{C_{2},\phi_{m}\}+
v^{n}\{C_{2},C_{n}\}\approx 0$, gives $u^{2}\approx -1/(2\mathcal{E})$. Finally, the consistency condition on $C_{1}$, given by
$\partial C_{1}/\partial \tau+\{C_{1},H_{C}\}+u^{m}\{C_{1},\phi_{m}\}+
v^{n}\{C_{1},C_{n}\}\approx 0$, provides $u^{1}\approx \Delta\mathcal{E}/2
- (\tilde{p}/\mathcal{E})(\Delta\mathcal{E}-(B/2)\tau)$, assuming
$q^{0}(\tau_{0})=q^{1}(\tau_{0})=\tau_{0}=0$.

The Dirac bracket of two functions $F$ and $G$ on the phase-space with coordinates $(q^{a},\chi,p_{b},\pi)$ can be calculated by the standard textbook definition,
thus providing $\{F,G\}^{*}=\{F,G\}+\{F,C_{1}\}\{\phi_{1},G\}-\{F,\phi_{1}\}\{C_{1},G\}-
[(1/2+\tilde{p}/\mathcal{E})q^{1}-(3/2)(\tilde{p}/\mathcal{E})q^{0}](\{F,C_{2}\}\{\phi_{1},G\}-\{F,\phi_{1}\}\{C_{2},G\})-[1/(2B\mathcal{E})](\{F,C_{2}\}\{\phi_{2},G\}-\{F,\phi_{2}\}\{C_{2},G\})-([3/(4\mathcal{E})]q^{0}-[1/(2\mathcal{E})]q^{1})(\{F,\phi_{1}\}\{\phi_{2},G\}-\{F,\phi_{2}\}\{\phi_{1},G\})$. Using this equation, it is straightforward to show that the Dirac bracket of
two observable quantities $O_{1}$ and $O_{2}$, which are functions on the 2d Minkowski space,
is given by
\begin{equation}\label{eq:diracbra}
\{O_{1},O_{2}\}^{*}=\frac{(\sqrt{-h})^{2}\epsilon^{ab}}{B}\frac{\partial O_{1}}{\partial q^{a}}\frac{\partial O_{2}}{\partial q^{b}}.
\end{equation}
In particular, it is now obvious that
$\{q^{a},q^{b}\}^{*}=(1/B)\epsilon^{ab}(\sqrt{-h})^{2}$. It follows that the 2d Minkowski
space is endowed with a symplectic form $\Omega_{(B)}=(B/2)\epsilon_{ab}\rmd q^{a}\wedge\rmd q^{b}$, which coincides with
the two form $\omega^{(2)}_{(B)}$ defining the central extension of Eq.~(\ref{eq:algpoincest})
(cf. Sec.~\ref{sec:1}). 

As we saw, the auxiliary gauge
degree of freedom $\chi$, which was introduced in order to neutralize the WZ term, is
associated with an extra primary first class constraint $\phi_{1}$, besides the usual one
$\phi_{2}$. As a consequence, we were led to assume the unusual canonical gauge conditions
$(C_{1},C_{2})$, which are ultimately responsible for the noncommutative structure that
the Dirac brackets above produce in 2d Minkowski space at the classical level. However, it is important to emphasize
that this noncommutativity is not an artifact of the gauge choice, since we consistently got
the same result in Ref.~\cite{artigo} by group theoretical methods, independently of any gauge
fixation as it was previously mentioned. 

Employing the Dirac brackets above, it is straightforward to check that the Hamilton
equations for the position coordinates can be expressed as $\dot{q}^{a}(\tau)=\{q^{a}(\tau),
H(q^{0},q^{1},\tau)\}^{*}$, where $H(q^{0},q^{1},\tau)=H(q(q^{0},q^{1}),E(q^{1}),\tau)=Bq^{1}-\dot{\mathcal{E}}(\tau)q^{0}$. Hence, the Hamiltonian $H(q^{0},q^{1},\tau)$ is a substitute for the vanishing canonical Hamiltonian $H_{C}=0$. 
Note that, in this gauge, the Hamiltonian does not coincide with the total energy, which is
given by $E_{T}=\mathcal{E}(\tau)+\mathcal{E}_{\mathrm{pot}}$, where
$\mathcal{E}_{\mathrm{pot}}=-E/2$ is the potential energy.

At the kinematical level, the standard canonical quantization of the system is well-defined.
In the non-canonical coordinates $q:=q^{0}-q^{1}$ and $p:=-Bq^{0}$ of the reduced phase-space, satisfying $\{q,p\}^{*}=-1$, it can be expressed in terms of the base kets $\{\vert x\rangle\}$ by  the Schr\"odinger representation
$\hat{q}:=x$, $\hat{p}:=-i\hbar(\partial/\partial x)$, and
$\hat{1}:=\mathbf{1}$, which maps the
Weyl--Heisenberg subalgebra of observables $\textrm{span}\{q,p,1\}$ onto the linear
space $\textrm{span}\{x,-i\hbar(\partial/\partial x),\mathbf{1}\}$ of unbounded Hermitian operators on the domain of the Schwartz space $\mathbf{\mathcal{S}}(\Re,\textrm{\textbf{C}})\subset L^{2}(\Re,dx)$ of rapidly decreasing smooth complex-valued functions. Although the algebra of observables cannot be fully quantized, the canonical quantization is enough
for quantizing all the observable quantities that are required to describe the quantum dynamics of the
system, such as position, momentum, potential energy, relativistic energy, or the
Hamiltonian (for further details see Ref.~\cite{artigo}).

In the coordinates $(q,p)$, the Hamiltonian operator splits into two parts
$\hat{H}(\hat{q},\hat{p},\tau)=\hat{H}_{0}(\hat{q},\hat{p})
+\hat{V}(\hat{p},\tau)$,
where $\hat{H}_{0}(\hat{q},\hat{p})=-B\hat{q}-\hat{p}$ and
$\hat{V}(\hat{p},\tau)=[\tilde{p}(\tau)/\sqrt{m^{2}+\tilde{p}(\tau)^{2}}]\hat{p}$. We remark
that the Hamiltonian is time-dependant only through the part $\hat{V}(\hat{p},\tau)$, so that
the interaction picture is the most suitable one for working the quantum dynamics out. Recall that
in this picture we use the eigenkets of the time-independent part of the
Hamiltonian as base kets. So, let us proceed by solving
the eigenvalue problem $\hat{H}_{0}\vert E\rangle=E\vert E\rangle$.

Performing the calculation, it is not difficult to see that
$\hat{H}_{0}$ has a continuous spectrum and that the normalized eigenfunctions are given by $\displaystyle
\langle x
\vert E\rangle=(1/\sqrt{2\pi\hbar})\exp[-(i/\hbar)(Ex+(B/2)x^{2})]$, so that $\langle E'\vert E\rangle=\delta(E'-E)$.
Note that classically
$H_{0}=Bq^{1}=-2\mathcal{E}_{\mathrm{pot}}(q^{1})$, so  
$\hat{H}_{0}(\hat{q},\hat{p})=-2\hat{\mathcal{E}}_{\mathrm{pot}}(\hat{q},\hat{p})$ has the meaning of a potential energy operator. Moreover, the total energy operator $\hat{\mathcal{H}}(\hat{q},\hat{p},\tau)=\mathcal{E}(\tau)
-\frac{1}{2}\hat{H}_{0}(\hat{q},\hat{p})$ satisfies $[\hat{\mathcal{H}},\hat{H}_{0}]=0$, therefore the eigenvectors of $\hat{H}_{0}$ are simultaneously total energy eigenstates. Hence, the eigenvalues
of the total energy operator are related with those of $\hat{H}_{0}$ through
$\hat{\mathcal{H}}(\tau)\vert E\rangle=E_{T}(\tau)\vert E\rangle$, where
$E_{T}(\tau)=\mathcal{E}(\tau)-E/2$.

In terms of the base kets
$\{\vert E\rangle\}$, the state ket of the system at $\tau=\tau_{0}$ is given  by $\vert\alpha
\rangle=\int^{\scriptscriptstyle +\infty}_{\scriptscriptstyle -\infty}\rmd E\, c_{E}(\tau_{0})\vert E\rangle$, where $c_{E}(\tau_{0})$ is some
known complex function of $E$ satisfying $\int^{\scriptscriptstyle +\infty}_{\scriptscriptstyle -\infty}\rmd E\vert c_{E}(\tau_{0})\vert^{2}=1$.
Then, in the non-canonical coordinates $(q,p)$, for $\tau>\tau_{0}$ the state ket in the Schr\"odinger picture will be given by $\vert\alpha,\tau_{0};\tau
\rangle=\int^{\scriptscriptstyle +\infty}_{\scriptscriptstyle -\infty}\rmd E\, c_{E}(\tau)e^{(iE/\hbar)\Delta\tau}\vert E\rangle$,
where the $c_{E}(\tau)$'s satisfy the basic coupled differential equations
$-i\hbar(\rmd c_{E}/\rmd\tau)=\int^{+\infty}_{-\infty}\rmd E'\langle E\vert \hat{V}\vert E'\rangle e^{[-i(E-E')/\hbar]
\Delta\tau}c_{E'}(\tau)$, which characterize the time evolution of a state ket in the interaction picture.

Finally, using $\langle E\vert \hat{V}\vert E'\rangle=(\tilde{p}/\mathcal{E})[-iB\hbar(\partial \delta(E'-E)/\partial E') -E'\delta(E'-E)]$, we can straightforwardly 
solve the resulting linear homogeneous partial differential equations by the method of
separation of variables to get
\begin{equation}\label{eq:coef}
c_{E}(\tau)=\exp\!\Bigg[\frac{\rmi(S(E,\tau)-E\Delta\tau)}{\hbar}\!\Bigg]
c_{E-\Delta\mathcal{E}}(\tau_{0}),
\end{equation}
where $S(E,\tau)$ is the action function given by Eq.~(\ref{eq:acfunc}).

\section{Noncommutative 2d Minkowski space}
\label{sec:3}
In this section, we will show that the anomaly-free 2d relativistic particle lives in a quantum
space-time. Then, we are going to build a Gaussian wave packet, in order to show that a
Planck length is well-defined in two dimensions. Finally, we will work out a dispersion
relation, so that we can discuss the physical aspects of the noncommutativity in connection
with the deformation of the Poincar\'e
symmetry which was introduced in Sec.~\ref{sec:1}.

The Hamiltonian analysis of the anomaly-free relativistic particle, which was
performed in the previous section, showed that its reduced phase--space is surprisingly
isomorphic to the 2d Minkowski space, for it is spanned by the
position coordinates $q^{a}$ alone. Now, exploiting this isomorphism, we can straightforwardly
apply Dirac's quantum condition to the fundamental Dirac brackets $\{q^{a},q^{b}\}^{*}$
that are defined throughout Minkowski space (cf.~Eq.~(\ref{eq:diracbra}) and
the discussion on the kinematical quantization of the anomaly-free relativistic particle in
Sec.~\ref{sec:2}),
which shows that the operators
corresponding to the components of the position two vector do not commute with each other:
\begin{equation}\label{eq:noncomm}
[\widehat{q^{a}},\widehat{q^{b}}]=-\frac{\rmi\hbar\epsilon^{ab}(\sqrt{-h})^{2}}{B}.
\end{equation}

Therefore, the non-trivial ($B\ne 0$) 2d relativistic elementary systems live in a
noncommutative 2d Minkowski space. This result is analogous to the noncommutativity which
is found in the configuration--space of an electron in an external constant magnetic field and
constrained to the lowest Landau level~\cite{jackiw6}. It is worth mentioning that the
anomaly-free 2d particle corresponds to a relativistic version of the Landau system, so that the extended Poincar\'e algebra $\bar{\textrm{\i}}^{1}_{2}$ is related to the
one-dimensional oscillator algebra os(1) --- which is the basic symmetry in the Landau problem --- 
by the Weyl unitary trick.

It is important to emphasize that Eq.~(\ref{eq:noncomm}) has not been assumed a priori. Ultimately, this result stems from the symplectic structure which the
non-trivial two cocycles $\xi_{B}(g',g)$ in $H^{2}_{0}(\mathcal{P},\Re)$ induce in 2d Minkowski space at the classical level (cf. Sec.\ref{sec:1}). Although the noncommutative quantum
anomaly-free 2d relativistic particle based in Eq.~(\ref{eq:noncomm}) may be formulated in
terms of a Groenewold--Moyal star product in space--time, in the present work we shall tie to
the operatorial approach that was adopted in the previous section, in which one works with
funtions of the noncommuting operators $\widehat{q^{a}}$ and with the ordinary product of
such functions.

Due to the Cauchy--Schwarz inequality, the commutation relations of Eq.~(\ref{eq:noncomm})
imply the uncertainty relation
$\langle (\Delta\widehat{q^{1}})^{2}\rangle\langle (\Delta\widehat{q^{0}})^{2}\rangle\ge
(1/4)|\langle[\widehat{q^{1}},\widehat{q^{0}}]\rangle|^{2}=\hbar^{2}/(4B^{2})$, where the dispersions of the coordinates are given by
$\langle (\Delta\widehat{q^{a}})^{2}\rangle=\langle (\widehat{q^{a}})^{2}\rangle-
\langle \widehat{q^{a}}\rangle^{2}$. Unlike the usual noncommutativity in space, this
uncertainty relation introduces a new kind of fuzziness of the points in 2d Minkowski space,
which precludes a simultaneous localization of the events in time and space. 
In order to probe this quantum space--time, we will now examine the time evolution of a
Gaussian wave packet.

In our formalism, the wave function for a relativistic particle state ket in the
Schr\"odinger picture at time $\tau$ is given in the $E$-representation (cf.~Sec.~\ref{sec:2})
by $\psi(E,\tau)=\langle E\vert\alpha,\tau_{0};\tau\rangle=
\exp[(\rmi/\hbar)E\Delta\tau]c_{E}(\tau)$. Thus, we can prepare the initial state of the
system as a Gaussian wave packet by setting 
$\psi(E,\tau_{0})=c_{E}(\tau_{0})=[1/(\pi^{1/4}\sqrt{d})]\exp[\rmi kE-(E-E')^{2}/(2d^{2})]$, which describes a plane wave with wave number $k$ propagating in $E$-space, modulated by a Gaussian profile
with width $d$ centered on $E=E'$. Note that the $E$-space is essentially identical with
space itself, since
$\widehat{q^{1}}=(1/B)\hat{H}_{0}$ so that the base kets $\{\vert E\rangle\}$ are also position
eigenkets $\widehat{q^{1}}\vert E\rangle=q^{1}\vert E\rangle$ with eigenvalues $q^{1}=E/B$.

At a later time $\tau>\tau_{0}$, the wave packet $\psi(E,\tau)$ will be given by the formula
above with $c_{E}(\tau)$ determined by Eq.~(\ref{eq:coef}). Moreover, it is easy to check that the wave packet is normalized to unit, $\langle\alpha,\tau_{0};\tau\vert\alpha,\tau_{0};\tau\rangle=
\int^{\scriptscriptstyle +\infty}_{\scriptscriptstyle -\infty}\rmd E\, |c_{E}(\tau)|^{2}=1$.
Thus, we can calculate the expectation value of any observable $\hat{A}$ in a wave packet
state at time $\tau$ simply by $\langle\hat{A}\rangle(\tau)=
\langle\alpha,\tau_{0};\tau\vert\hat{A}\vert\alpha,\tau_{0};\tau\rangle=
\int^{\scriptscriptstyle +\infty}_{\scriptscriptstyle -\infty}\rmd E\int^{\scriptscriptstyle +\infty}_{\scriptscriptstyle -\infty}\rmd E'c_{E'}^{*}c_{E}\exp[(\rmi/\hbar)(E-E')\Delta\tau]
\langle E'\vert\hat{A}\vert E\rangle$.

Applying this equation, it is not difficult to show that
$\langle\widehat{q^{1}}\rangle(\tau)=\langle\widehat{q^{1}}\rangle(\tau_{0})+
\Delta\mathcal{E}/B$, which consistently coincides with the Hamilton equation describing the
uniformly accelerated relativistic motion of the particle at the classical level (cf.~Sec.~\ref{sec:2}). Further we
can straightforwardly calculate the dispersion of the position coordinate
$\langle (\Delta\widehat{q^{1}})^{2}\rangle(\tau)=(1/B^{2})\langle (\Delta\hat{H}_{0})^{2}
\rangle(\tau)=d^{2}/(2B^{2})$, which in particular shows that the parameter $d$ measures the
spread of the wave packet on space indeed.

In order to calculate the expectation value of the time coordinate, we first note that
$\langle\widehat{q^{0}}\rangle(\tau)=-(1/B)\langle\hat{p}\rangle(\tau)$ and $\langle\hat{p}\rangle(\tau)=-B\langle\hat{q}\rangle(\tau)-\langle\hat{H}_{0}\rangle(\tau)$. It is not difficult to
show that $\langle\hat{H}_{0}\rangle(\tau)=E'+\Delta\mathcal{E}(\tau)$, which in particular
shows that $\langle\widehat{q^{1}}\rangle(\tau_{0})=(1/B)\langle\hat{H}_{0}\rangle(\tau_{0})=
E'/B$. Then,
with the aid of $\langle E'\vert\hat{q}\vert E\rangle=\rmi\hbar(\partial\delta(E-E')/
\partial E)$, we can straightforwardly compute $\langle\hat{q}\rangle(\tau)=\hbar k-\Delta\mathcal{E}/B+\Delta\tau$. Thus, $\langle\widehat{q^{0}}\rangle(\tau)=\hbar k+
\Delta\tau+E'/B$, which also provides an interpretation for the wave number $k=(1/\hbar)
[\langle\widehat{q^{0}}\rangle(\tau_{0})-\langle\widehat{q^{1}}\rangle(\tau_{0})]$. Finally,
choosing $\langle\widehat{q^{0}}\rangle(\tau_{0})=\tau_{0}$ we get $\langle\widehat{q^{0}}\rangle(\tau)=\tau$, which consistently coincides with the Hamilton equation for the time
coordinate at the classical level (cf.~Sec.~\ref{sec:2}). Note that with no loss of generality
we can assume 
$\langle \widehat{q^{1}}\rangle(\tau_{0})=\langle\widehat{q^{0}}\rangle(\tau_{0})=\tau_{0}=0$,
which amounts to setting $k=E'=\tau_{0}=0$.

The computation of the dispersion of the time coordinate is more involved. First we note that
$\langle (\Delta\widehat{q^{0}})^{2}\rangle(\tau)=(1/B^{2})\langle (\Delta\hat{p})^{2}
\rangle(\tau)=(1/B^{2})[\langle \hat{p}^{2}\rangle-\langle \hat{p}\rangle^{2}]$ and
$\langle\hat{p}^{2}\rangle=\langle(-B\hat{q}-\hat{H}_{0})^{2}\rangle=B^{2}\langle\hat{q}^{2}\rangle+\langle\hat{H}_{0}^{2}\rangle+B\langle\hat{q}\hat{H}_{0}\rangle+B\langle\hat{H}_{0}\hat{q}\rangle$. Then, after a
tedious calculation we get $\langle\hat{q}^{2}\rangle=\hbar^{2}/(2d^{2})+(\hbar k-
\Delta\mathcal{E}/B+\Delta\tau)^{2}$, $\langle\hat{H}_{0}^{2}\rangle=d^{2}/2+(E'+\Delta\mathcal{E})^{2}$, $\langle\hat{q}\hat{H}_{0}\rangle=-(\rmi\hbar)/2+(E'+\Delta\mathcal{E})(\hbar k-
\Delta\mathcal{E}/B+\Delta\tau)$, and $\langle\hat{H}_{0}\hat{q}\rangle=(\rmi\hbar)/2+(E'+\Delta\mathcal{E})(\hbar k-
\Delta\mathcal{E}/B+\Delta\tau)$. It follows that $\langle (\Delta\widehat{q^{0}})^{2}\rangle(\tau)=\hbar^{2}/(2 d^{2})+d^{2}/(2B^{2})$.

The first remarkable fact that we learn from the calculations above is that the width of the
wave packet does not increase with time, since the dispersions
$\langle (\Delta\widehat{q^{a}})^{2}\rangle(\tau)$ are constant.
Not less surprising is the fact that the dispersion of the time coordinate attains to a minimum
at $d=\sqrt{|B|\hbar}$, which corresponds to the root mean square deviations
$\sqrt{\langle (\Delta\widehat{q^{0}})^{2}\rangle}=\sqrt{\hbar/|B|}$ and $\sqrt{\langle (\Delta\widehat{q^{1}})^{2}\rangle}=(\sqrt{2}/2)\sqrt{\hbar/|B|}$. Although such a minimum temporal 
dispersion Gaussian wave packet is not an
exact coherent state, since its uncertainty product is only approximately minimum (for a large
$|B|$), $\langle (\Delta\widehat{q^{1}})^{2}\rangle\langle (\Delta\widehat{q^{0}})^{2}\rangle
=\hbar^{2}/(2B^{2})\approx\hbar^{2}/(4B^{2})$, it is the Gaussian wave packet that more closely resembles a classical relativistic particle or a space--time point. Hence, we can define a 2d Planck length $l_{P}$ in terms of
the shortest observable time interval, given by the minimum $\sqrt{\langle (\Delta\widehat{q^{0}})^{2}\rangle}$: 
\begin{equation}\label{eq:planck}
l_{P}=\sqrt{\frac{\hbar}{|B|}}.
\end{equation}

It is worth mentioning that this noncommutative space--time does not present any discrete structure
in the absence of curvature, since the position coordinates $q^{a}$ have been quantized
by unbounded operators with continuous spectra (cf. Sec.~\ref{sec:2}). Moreover, although the time
coordinate $q^{0}$ is involved in the commutation relations of Eq.~(\ref{eq:noncomm}), the
particle theory defined by Eq.~(\ref{eq:acaoiltonia2}) is perfectly unitary in the sense
of Ref.~\cite{gomis},
due to Eq.~(\ref{eq:coef}). Nevertheless, causality violations are expected at Planck scale,
since the quantum description of the system employs a time operator. 

To conclude this section, we will discuss below some physical aspects of the noncommutative space-time introduced above which are connected to the deformation of the Poincar\'e algebra given by Eq.~(\ref{eq:algpoincest}). First recall that the central charge is a free parameter that must be fixed at the outset, since it is characteristic of the 2d relativistic elementary systems (cf.~Sec.~\ref{sec:1}). Further, since the beginning of Sec.~\ref{sec:2} we have been consistently assuming that $B\ne 0$, for we are concerned with the physical interpretation of the non-trivial elementary systems. We have already pointed out that the non-trivial elementary systems are a peculiarity of the two dimensional case and that they exhibit properties which are totally different from those of the trivial ones.

If we assume that the central charge is exactly zero, then the description of the 2d relativistic particle proceeds in the standard way  and no noncommutativity shows up. In this case, the system is invariant under the 2d Poincar\'e group and the quantization is performed in the phase-space, $[\widehat{q^{1}},\widehat{p_{1}}]=-\rmi\hbar$, so that the deformation parameter is the Planck constant. However, for $B\ne 0$ the anomaly-free 2d relativistic particle is invariant under the extended Poincar\'e group and we have shown in Sec.~\ref{sec:2} that the quantization of this system is performed in space-time. In this case, we can see from Eq.~(\ref{eq:noncomm}) that the deformation parameter is the central charge $B$, which is independent of the Planck constant. Note that the 2d Minkowski space becomes commutative in the strong field limit $B\to\infty$, which corresponds to the low energy limit $l_{P}\to 0$ due to Eq.~(\ref{eq:planck}). 

On the other hand, we cannot recover the standard Poincar\'e invariant theories from the non-trivial elementary systems, since the latter cannot be continuously deformed into the former. In fact, the anomaly-free relativistic particle is not well-defined in the undeformed case $B\to 0$ --- neither is the corresponding irrep (cf. Sec. III-A of Ref.~\cite{artigo}) ---, since then the symplectic form $\Omega_{(B)}$ defining Eq.~(\ref{eq:noncomm}) is not invertible. But this limitation can be alleviated by considering that the limit $B\to 0$ corresponds to sub-Planckian length scales (see Eq.~(\ref{eq:planck})), where all physical theories are expected to break down anyway. Therefore, we will be concerned about the strong field limit $B\to\infty$ below, since only the low energy limit is relevant from the physical point of view.

Relativistic theories based on deformed Poincar\'e algebras with two
relativistic-invariant scales (a velocity scale $c$ and a length scale $l_{P}$) are sometimes generically called doubly-special relativity
theories (DSR's)~\cite{camelia2}. In this context, an interesting proposal by Magueijo and Smolin (MS)~\cite{smolin} introduced the so called $\kappa$-deformations of the Poincar\'e algebra in four dimensions, which
have the merit of preserving the Lorentz invariance by assuming that the Lorentz subgroup of transformations acts non-linearly on momentum space.  In the sense of introducing a second observer-independent scale, through Eq.~(\ref{eq:planck}), the deformation of the 2d Poincar\'e algebra of Eq.~(\ref{eq:algpoincest}) is suitable for describing a DSR framework. However, it must be emphasized that 
the extended Poincar\'e algebra $\bar{\textrm{\i}}^{1}_{2}$ is not $\kappa$-deformed in the sense of MS, since its deformation consists in the introduction of a non-trivial center in the sector of translations of the 2d Poincar\'e algebra without modifying the generator of boosts, which is kept acting linearly on momentum space.

In the context of deformed Poincar\'e algebras, in general we expect that corrections to the standard laws of relativistic mechanics apply at Planck scale. So, we will now show that the modifications which the extended Poincar\'e algebra of Eq.~(\ref{eq:algpoincest}) produces to the kinematical laws --- parametrized by the observer-independent deformation parameter $B$ --- do not violate the 2d relativity principle. Further, we will test wether such modifications eventually vanish at low energies. First note that commutation relations like Eq.~(\ref{eq:noncomm}) usually indicate Lorentz symmetry
breaking. Generically speaking, whenever this happens the energy-momentum transformation rules remain unmodified, despite the Planck scale modification of the dispersion
relations. On the other hand, when the
transformation rules between inertial observers are deformed, the laws of energy-momentum
conservation must also be necessarily modified so that they have the same structure for all
inertial observers.

Recall that a dispersion relation denotes the functional dependence between the relativistic energy and the kinematical momentum of a free particle, which is set up by the mass-shell condition. In the case of the extended Poincar\'e algebra, it should be emphasized that the anomaly-free relativistic particle is a non-conservative system which receives an energy flux from the interaction with an external electric-like force field $B$ (cf. Sec.~\ref{sec:2}). It follows that the bivectors associated with the canonical momentum $p_{a}$, the kinematical momentum $\tilde{p}_{a}=p_{a}+(B/2)\epsilon_{ab}q^{b}$, and the energy-momentum $\mathcal{N}_{a}=p_{a}-(B/2)\epsilon_{ab}q^{b}$ are all distinct. Further, the mass-shell condition can be expressed in terms of the kinematical momentum bivector by the constraint $\phi_{2}=\tilde{p}^{a}\tilde{p}_{a}-m^{2}$, the gauge of which is fixed by the canonical gauge condition $C_{2}=\tilde{p}(\tau)-\tilde{p}(\tau_{0})-B(\tau-\tau_{0})$, where $\tilde{p}(\tau):= \tilde{p}_{1}(\tau)$ is the kinematical momentum.

At the classical level, the model is completely specified in the reduced phase-space, with coordinates $\{q^{a}\}$, by the following Hamiltonian system:
\begin{equation}\label{eq:hamsys}
\fl \tilde{p}^{a}\tilde{p}_{a}=m^{2},\quad\dot{\tilde{p}}_{1}=B,\quad
\{q^{a},q^{b}\}^{*}=\frac{\epsilon^{ab}(\sqrt{-h})^{2}}{B},\textrm{ and}\quad H(q^{a},\tau)=\dot{\tilde{p}}_{a}q^{a},
\end{equation}
where the dot stands for $\rmd/\rmd\tau$. 
As a result, the deformed dispersion relation is given by $\mathcal{E}(\tilde{p}_{1}):=-\tilde{p}_{0}=\sqrt{m^{2}\!  +\!  (\tilde{p}_{1})^{2}}$. It can be seen that the deformation does not modify the functional dependence between $\mathcal{E}$ and $\tilde{p}_{1}$, with respect to the standard case; it merely introduces a time dependence into the relativistic energy $\mathcal{E}(\tau)=\mathcal{E}(\tilde{p}(\tau))$ --- where $\tilde{p}(\tau)=\tilde{p}(\tau_{0})+B(\tau-\tau_{0})$ ---, implicitly through $\tilde{p}_{1}(\tau)$ when the canonical gauge condition $C_{2}$ is imposed. Note that  $\mathcal{E}(\tau)$ is distinct from the total energy $E_{T}(\tau)=-p_{0}(\tau)=\mathcal{E}(\tau)-(B/2)q^{1}(\tau)$, which includes the potential energy of the particle.

Using the definition of the Dirac brackets given by Eq.~(\ref{eq:diracbra}), it is not difficult to check that the Hamiltonian system of Eq.~(\ref{eq:hamsys}) is manifestly invariant under the rigid extended Poincar\'e transformations $\tilde{p}_{a}\to \tilde{p}_{b}(\Lambda^{-1})^{b}_{\verb+ +a}$ and $q^{a}\to \theta^{a}+\Lambda^{a}_{\verb+ +b}q^{b}$. Indeed, the result follows after discarding a total derivative with respect to $\tau$ in the transformed Hamiltonian and taking into account the fact that in any inertial frame it is always possible to fix the parametrization of the world line so that the particle feels a constant physical force equal to $B$ in its rest frame (i.e. the equation $\dot{\tilde{p}}_{1}=B$ can be set up in all inertial frames).
Hence, the deformed dispersion relation has the same structure in all inertial frames, which means that there is
no preferred class of observers as long as we formulate the 2d special relativity principle in terms of the extended Poincar\'e group.

Before we briefly discuss the low energy limit, let us mention that when $B\ne 0$ the extended Poincar\'e algebra reduces to the Weyl-Heisenberg subalgebra. It is easier to see this fact by considering the Poisson bracket realization of the Eq.~(\ref{eq:algpoincest}) in the Hamiltonian system of Eq.~(\ref{eq:hamsys}), which is provided by the Hamiltonians (i.e. the generators of the canonical transformations) given by $u_{a}=Bq^{b}\epsilon_{ba}$, $u_{2}=(m^{2}/2B)+[B/(2\sqrt{-h})]q_{a}q^{a}$, and $u_{3}=-1$, which satisfy
\begin{eqnarray}\label{eq:poissonrealiz}
\{u_{a},u_{2}\}^{*} = \sqrt{-h}\epsilon^{\verb+ +b}_{a}u_{b},\quad
\{u_{a},u_{b}\}^{*}=B\epsilon_{ab} u_{3},\textrm{ and}& &\nonumber\\
\{ u_{a},u_{3}\}^{*}  =  \{ u_{2},u_{3}\}^{*}=0.& &
\end{eqnarray}
Since the identity $u_{a}u^{a}-2(B/\sqrt{-h})u_{2}u_{3}=m^{2}/\sqrt{-h}$ holds, for $B\ne 0$ we can eliminate $u_{2}$ in terms of $\{u_{a},u_{3}\}$, which close the Weyl-Heisenberg subalgebra of Eq.~(\ref{eq:poissonrealiz}). Hence, all the extended Poincar\'e symmetry of the Hamiltonian system of Eq.~(\ref{eq:hamsys}) reduces to this subalgebra, which is dual to the algebra of the fundamental dynamical variables $\{q^{a},1\}$ itself (for a more formal treatment of this result cf.~Ref.~\cite{artigo}).  

The analysis of the behaviour of the Hamiltonian system of Eq.~(\ref{eq:hamsys}) in the low energy limit is problematic, since it involves the manipulation of several quantities that go to infinity in the strong field limit $B\to\infty$. Indeed, although the extended Poincar\'e symmetry is well-defined in the strong field limit, in the sense that the 2d Minkowski space becomes commutative (cf. the discussion above), the Hamiltonian $H(q^{a},\tau)$ is not so. Nevertheless, in a loose manner we can say that the Hamiltonian behaves asymptotically as $H(q^{a},\tau)=Bq^{1}-Bq^{0}$, so that the Hamilton equations $\dot{q}^{a}=\{q^{a},H\}$ imply that $q^{0}=q^{1}$. This suggests that in the low energy limit all particles move at the speed of light. In order to make this argument more rigorous, first observe that the effect of the strong field limit is to produce an infinite proper acceleration $\omega_{0}=B/m$
on the particle. Thus, we can avoid the singular limit $B\to\infty$ by taking the massless limit $m\to 0$, which produces the same effect with the same asymptotic Hamiltonian $H(q^{a},\tau)=Bq^{1}-Bq^{0}$, now well-defined whith a finite central charge.

We have shown above that the deformation of the Poincar\'e algebra given by Eq.~(\ref{eq:algpoincest}) produces no modifications to the standard laws of relativistic mechanics. The effect of the deformation is to produce an interaction with an external field that drives the particles into an uniformly accelerated motion. In the low energy limit, the relativistic energy of all the particles is dominated by the external field, so that they behave as massless particles.

To conclude, we remark that the problematic nature of the limits $B\to 0$ and $B\to\infty$ implies that the non-trivial elementary systems are well-defined only for a finite value of $B$. Thus, the central charge is better interpreted as a coupling constant rather than a deformation parameter, so that it will be consistently related to a cosmological constant in the next section. Summarizing, in this section we have succeeded in providing a physical interpretation for a noncommutative 2d Minkowski space in terms of a non-conservative interacting Hamiltonian system, formulated in the framework of standard relativistic mechanics.

\section{Extended dilaton gravity theories}
\label{sec:4}
In Sec.~\ref{sec:1} we saw that the determination of the Hilbert space of any
type of elementary system at the quantum level depends on the formulation of the 2d special
relativity principle in terms of some fixed $\bar{\mathcal{P}}_{(B)}$. In the previous section, we learned that 
the deformation of the 2d Poincar\'e symmetry sets an interaction that introduces a noncommutative structure in the 2d Minkowski space, without modifying the standard laws of relativistic mechanics. Moreover, the central charge $B$ acts as a coupling constant that defines a 2d Planck length $l_{P}$.

We have seen that this peculiar interaction is similar to a constant
electric force field $\omega^{(2)}_{(B)}$ defined throughout the 2d Minkowski space,
the potential one form of which $\beta^{(1)}_{(B)}$ is a WZ potential that drives all
particles into an uniformly accelerated relativistic motion. It turned out that such
interacting systems in the non-trivial case ($B\ne 0$) constitute new kinds of elementary
systems in their own right, the anomaly-free description of which necessarily requires the
correction of the
2d relativity principle formulated above. In Sec.~\ref{sec:2} we worked out
the reduced phase-space of the anomaly-free 2d relativistic particle, in order to show that
the non-trivial elementary systems live in a noncommutative 2d Minkowski space.
Finally, in the previous section we studied the physical aspects of this noncommutativity in
the absence of gravity.

Now we shall proceed, aiming at providing a gravitational interpretation for this
noncommutative structure. We point out that the noncommutativity is produced by the electric-like constant force field $\omega^{(2)}_{(B)}$, which endows the 2d Minkowski
space with a symplectic form. On the other hand, considering the universal character of such interaction, it is natural to regard it as a property of space-time --- just like
the metric ---, rather than of the particles themselves. Hence, we will extend the 2d generalized dilaton gravity theories by an additional U(1) sector that gauges the extra symmetry
associated with the center of the extended Poincar\'e algebra. The hope is to generalize the WZ
potential --- which was treated as an externally applied field in the previous sections --- to
a dynamical field which should locally determine the noncommutative structure of the curved space-time.

To accomplish such an extension, we are proposing a non-standard dilaton--Maxwell theory,
which allows for a non-minimal coupling between the extra U(1) sector and the geometric one.
It is non-standard in the sense that the additional Lagrange multiplier that we are going to introduce in the Maxwell sector will appear
linearly in the potential term of the action, rather than quadratically. The idea is to
set a covariantly constant field strength throughout space-time, in order to
generalize the electric-like force field $\omega^{(2)}_{(B)}$ to the case with curvature. Note that
the associated gauge potential will still be called a WZ potential because it is intended
to be a generalization of $\beta^{(1)}_{(B)}$ --- which satisfies the WZ consistency condition of Eq.~(\ref{eq:wesszumino}) ---, in spite of the fact that it is an ordinary U(1) gauge potential. In
this regard, it is important to emphasize that the extra BF term that we are going to introduce in
the dilaton models should not be confused with a WZ term in the same sense that is employed in the context of
anomalous gauge theories.  

Employing Cartan variables, the action can be written in first order as
\begin{equation}\label{eq:dilest}
\fl
\!\! S^{\mathrm{(EFOG)}}\!=\frac{1}{G}\!\int_{M}\!\!\left[ \eta_{a}De^{a}\!+\eta_{2}\rmd\omega+G\eta_{3}\rmd a+\epsilon\!\left(\textrm{{\LARGE
$\nu$}}(\eta^{a}\eta_{a},\eta_{2})+\frac{BG\eta_{3}}{W(\eta_{2})}\right)\right],
\end{equation}
where $M$ is the space--time manifold, $G$ is the 2d gravitational constant of dimension
$[G]=[\hbar]^{-1}$, $\epsilon=(\epsilon_{ab}e^{a}e^{b})/2$ is the
volume two form, and $De^{a}=\rmd e^{a}+\sqrt{-h}\epsilon^{a}_{\verb+ +b}\omega e^{b}$. The zweibeine $e^{a}$ and
the dilaton $\eta_{2}$ are dimensionless, the spin connection $\omega$ has dimension
$[\omega]=L^{-1}$, and
the Lagrange multipliers $\eta_{a}$ (related to the torsion of space--time) have dimension
$[\eta_{a}]=L^{-1}$. Note that our conventions are such that
$e=\mathrm{det}e^{a}_{\mu}=-\sqrt{-g}$, so that the usual textbook definitions for the Hodge-$\ast$ operation and the Ricci-scalar imply $\ast\epsilon=1$ and $R=-2\ast\!\rmd\omega$. In addition,
the torsionless part of the spin connection is given by $\tilde{\omega}=-e_{a}\ast\!\rmd e^{a}$
and $R>0$ for $\mathrm{dS}_{2}$.  

The extension introduces the real WZ potential $a$ of dimension $[a]=L^{-1}\times[\hbar]$,
which generalizes $\beta^{(1)}_{(B)}$ and is responsible for the aforementioned interaction
that characterizes the non-trivial elementary systems, and a fourth dimensionless
Lagrange multiplier $\eta_{3}$ that describes a dynamical U(1) charge. Provided there are no matter
couplings to $a$, the field equation resulting from its variation implies the conservation of
$\eta_{3}$. It is
also introduced a non-vanishing free parameter $B$ of dimension $[B]=L^{-2}\times[\hbar]$,
which sets the intensity of the covariantly constant electric-like force described previously
and shall be interpreted as a new fundamental physical constant that determines a 2d Planck length $l_{P}$, according to Eq.~(\ref{eq:planck}). Note that $B$ is not absorbed into $\eta_{3}$
because if we did so it would arise as a constant of motion and we do not expect
physical states to have a central charge spectrum, due to the Bargmann superselection rule
(cf. Sec.~\ref{sec:1}).

The dynamical content of the theory is determined by the Lorentz invariant functions
$W(\eta_{2})$ and {\LARGE $\nu$}$(\eta^{a}\eta_{a},\eta_{2})$. The former is a non-vanishing function of the dilaton that allows a non-minimal
coupling between the U(1) gauge sector and the geometric one --- such that the
minimal coupling corresponds to a constant $W(\eta_{2})$ --- and the latter splits into
{\LARGE $\nu$}$(\eta^{a}\eta_{a},\eta_{2})=
 U(\eta_{2})[(\sqrt{-h}\,\eta^{a}\eta_{a})/2] + V(\eta_{2})$. Thus, the extended dilaton
theories of Eq.~(\ref{eq:dilest}) are characterized by three potentials: the usual
$U(\eta_{2})$ and $V(\eta_{2})$ potentials, together with an extra one $W(\eta_{2})$. 

These potentials allow the integration of two Casimir functions (constants of motion), which parametrize all the solutions for some specific model. The first one has the meaning of space--time energy and is given by $C_{1}=[(\sqrt{-h}\,\eta^{a}\eta_{a})/2]\rme^{-Q(\eta_{2})}- \bar{Z}(\eta_{2},\eta_{3})$, where
$Q(\eta_{2})=\int^{\eta_{2}}U(y)\rmd y$,
$\bar{Z}(\eta_{2},\eta_{3})=\int^{\eta_{2}}\rme^{-Q(y)}\bar{V}(y,\eta_{3})\rmd y$, and
$\bar{V}(\eta_{2},\eta_{3})=V(\eta_{2})+[BG\eta_{3}]/W(\eta_{2})$. The other is
$C_{2}=\eta_{3}$, which has already been interpreted above.  

We emphasize that the extended generalized theories (EFOG's) are not trivially
equivalent to the ordinary dilaton gravity models, since the U(1) gauge sector does
not decouple completely. In fact, it is straightforward to show that, in the absence
of matter, the field equations derived from Eq.~(\ref{eq:dilest}) encompass those of the unextended theories with
{\LARGE $\bar{\nu}$}$(\eta^{a}\eta_{a},\eta_{2};\eta_{3}^{0}BG)=U(\eta_{2})[(\sqrt{-h}\,\eta^{a}\eta_{a})/2] + \bar{V}(\eta_{2},\eta_{3}^{0})$, together with the topological equation
\begin{equation}\label{eq:top}
\rmd a+\frac{B\,\epsilon}{W(\eta_{2})}=0,
\end{equation}
where $\eta_{3}^{0}$ denotes the on-shell value of the U(1) charge.

Concerning the field equations derived from Eq.~(\ref{eq:dilest}), it is important to note
that in general the WZ potential $a$
will not be redundantly expressed in terms of the geometric variables $(e^{a},\omega)$~\cite{cangemi2}, but
its dual field strength $\ast \rmd a=-B/W(\eta_{2})$ will be determined by the
dilaton. Moreover, although Eq.~(\ref{eq:top})
can always be integrated locally, it will generally require an atlas containing more than one
chart for the space--time manifold $M$, due to the singularities of the $a$
field~\cite{jackiw2}.

Indeed, if $M$ is topologically trivial it can be contracted to a bounded manifold, so
that the dilaton is kept everywhere non-singular and the integral over $M$ of the second term on the left side of Eq.~(\ref{eq:top}) is made a
well-defined finite non-vanishing quantity. However, we come to a contradiction if $a$ is non-singular, since then it must be single-valued and the first term on the left side of Eq.~(\ref{eq:top}) integrates to zero.

We are showing below that the extended dilaton theories provide a high energy correction to the unextended ones, which produces a noncommutativity in curved 2d space-time. However, we will only prove that this noncommutativity is described by a Groenewold--Moyal star product in the low energy limit. For the high energy case we will merely provide a plausible conjecture that the noncommutativity is described by a star product which is not of the Groenewold--Moyal type, since a rigorous proof of this conjecture is out of the scope of the present work.   

It is known that the Eddington-Finkelstein (EF) gauge is the most natural one for solving the field equations of 2D dilaton gravity~\cite{grumiller1}. In this gauge, a redefinition of the dilaton $\tilde{\eta}_{2}$ is taken as a coordinate, while it is usually assumed that $\eta_{2}$ is positive and finite inside a patch. Whenever black holes are present, curvature singularities are typically located at the boundaries of the range for $\tilde{\eta}_{2}$.

In EF coordinates, it is not difficult to see that the coefficient of $B$ in Eq.~(\ref{eq:top}) is finite inside a patch, so that it can be ignored in the strong field limit $B\to\infty$. Thus, when $B$ is very large the Eq.~(\ref{eq:top}) can be locally approximated by
$\rmd a\approx -(B/2)\tilde{\epsilon}_{\mu\nu}\rmd x^{\mu}\wedge\rmd x^{\nu}$ (where $\tilde{\epsilon}_{\mu\nu}=\tilde{\epsilon}^{\mu\nu}$ is the Levi--Civita symbol with $\tilde{\epsilon}_{01}=1$), which implies
that $a\approx -B\tilde{\epsilon}_{\mu\nu}x^{\mu}\rmd x^{\nu}$.
Moreover, by a similar argument the last term in Eq.~(\ref{eq:dilest}) behaves asymptotically as a constant
$\int_{M}\!\epsilon B[\eta_{3}/W(\eta_{2})]\approx
\int_{M}\rmd x^{0}\rmd x^{1}B$, so that it can be dropped.
Hence, the U(1) sector completely decouples from the geometric one and can be integrated out, with the result that the ordinary dilaton theories are recovered in the low energy limit. It follows that the extended models provide a high energy correction to the usual dilaton theories.

If we couple a test particle to the extended dilaton theories of Eq.~(\ref{eq:dilest}),
$S_{P}=\int\rmd\tau[m\sqrt{g_{\mu\nu}\dot{x}^{\mu}\dot{x}^{\nu}}+a_{\mu}\dot{x}^{\mu}]$, it
is not difficult to see that it decouples in the low energy limit
$B\to\infty$, since then we can neglet the mass term. Thus, using the approximation for the WZ
potential given above, we get $S_{P}\approx -\int\rmd\tau B\tilde{\epsilon}_{\mu\nu}x^{\mu}
\dot{x}^{\nu}$, which shows that the variables $x^{\mu}$ and $p_{\nu}=-B
\tilde{\epsilon}_{\mu\nu}x^{\mu}$ are canonically conjugate $\{x^{\mu},p_{\nu}\}=
\delta^{\mu}_{\nu}$, so that $\{x^{\mu},x^{\nu}\}=(1/B)\tilde{\epsilon}^{\mu\nu}$
(compare with Eq.~(\ref{eq:noncomm})). Therefore, the extended dilaton theories predict a very small noncommutativity in curved space-time at low energies, which is described by a Groenewold--Moyal star product given by $x^{\mu}\star x^{\nu}-x^{\nu}\star x^{\mu}=-\rmi(l_{P})^{2}\tilde{\epsilon}^{\mu\nu}$ (for $B>0$).

In the light of the discussion above, we can see that Eq.~(\ref{eq:top}) provides precisely
the generalization of the electric-like force field $\omega^{(2)}_{(B)}$ that we were looking
for. By analogy to the 2d Minkowski space case, we now conjecture that the field strength $da=
-[B/(2W)]\epsilon_{ab}\rme^{a}\wedge\rme^{b}$ endows the curved space-time manifold with a noncommutative structure. At the classical level, this noncommutativity should be described by the contravariant tensor field which is dual to $da$ and defines a bracket of two functions on space-time given by
$\{F,G\}:=-(W/B)\epsilon^{ab}\rme_{a}(F)\rme_{b}(G)$, or $\{F,G\}:=[W(\eta_{2})/(B\sqrt{-g})]
\tilde{\epsilon}^{\mu\nu}(\partial F/\partial x^{\mu})(\partial G/\partial x^{\nu})$ in local coordinates. This
means that the star product in curved space-time must be given by
\begin{equation}\label{eq:star}
\fl
x^{\mu}\star x^{\nu}-x^{\nu}\star x^{\mu}:=-\frac{\rmi\hbar
W(\eta_{2})\tilde{\epsilon}^{\mu\nu}}{B\sqrt{-g}}=-\rmi(l_{P})^{2}\frac{W(\eta_{2})\tilde{\epsilon}^{\mu\nu}}{\sqrt{-g}}\quad\textrm{(for $B>0$)},
\end{equation}
where an ordering prescription (such as the one of Weyl) is assumed for the right hand side
of Eq.~(\ref{eq:star}). Note that this star product is not of the Groenewold--Moyal type, but it is asymptotically equivalent to one of this type in the low energy limit $l_{P}\to 0$. However, we will leave a rigorous proof of Eq.~(\ref{eq:star}) at high energies
for a future work.

\begin{table}[!hbp]
\footnotesize\renewcommand{\arraystretch}{1.3}
\addtolength{\tabcolsep}{-5pt}
\caption{\label{tab:selec}\emph{Up-to-date list of extended models} - The parameters ($b$, $\lambda$, $E$)
define scale factors, ($\mu$, $\lambda_{0}$) are cosmological constants, ($c$, $\Lambda$)
denote constant curvature values, and $\phi_{1}$ is the initial ``constant dilaton vacuum''
point. Finally, ($\alpha$, $\beta$, $A$, $B$, $M$) are coupling constants.}
\begin{center}
\begin{tabular}{|l|c|c|c|c|}
\hline
\rowcolor[gray]{.9}[.95\tabcolsep]\textbf{Extended Model} & $\bf{U(\eta_{2})}$ & $\bf{V(\eta_{2})}$ & $\bf{W(\eta_{2})}$ & $\bf{BG}$\\
\hline
& & & &\\[-10pt]
1. Schwarzschild {\scriptsize $(\mathrm{D}>3,$}& $-\frac{D-3}{(D-2)\eta_{2}}$ & $0$ &
$-2(\eta_{2})^{-\frac{D-4}{D-2}}$ &  $\lambda_{D}^{2}$\\[-3pt]
{\scriptsize $\lambda_{D}^{2}=\lambda^{2}(D-2)(D-3))$} & & & &\\
2. Witten BH / CGHS & $-(\eta_{2})^{-1}$ & $0$ & $-2(\eta_{2})^{-1}$ & $4b^{2}$\\
3. Conf. Transf. CGHS & $0$ & $0$ & $-2$ & $4b^{2}$\\
4. Jackiw--Teitelboim  & $0$ & $0$ & $-2(\eta_{2})^{-1}$ & $-\Lambda$ \\
5. $\mathrm{(A)dS}_{2}$ Ground State & $-a(\eta_{2})^{-1}$ & $0$ & $-2(\eta_{2})^{-1}$ & $E$\\
6. Rindler Ground State & $-a(\eta_{2})^{-1}$ & $0$ & $-2(\eta_{2})^{-a}$ & $E$\\
7. BH Attractor & $0$ & $0$ & $-2\eta_{2}$ & $E$\\
\hline
8. All above: ab--family & $-a(\eta_{2})^{-1}$ & $0$ & $-2(\eta_{2})^{-(a+b)}$ & $E$\\
\hline
9. Katanaev--Volovich  & $\alpha$ & $\frac{\beta}{2}(\eta_{2})^{2}$ & $-2$ & $2\Lambda$ \\
10. de Sitter Gravity & $0$ & $\frac{\Lambda}{2}\eta_{2}$ & $-1$ & $\lambda_{0}$\\
11. Liouville Gravity & $a$ & $0$ & $\rme^{-\alpha\eta_{2}}$ & $\mu$\\
12. Scatt. Trivial Boson & generic & $UV+V'=0$ & $K\rme^{\int^{\eta_{2}}U(X)\rmd X}$ & \\[-3pt]
{\scriptsize $(K\ne 0)$} & & & &\\
13. Scatt. Trivial Fermion & $0$ & $0$ & $1$ & const.\\
14. Reissner--Nordstr\"om & $-\frac{1}{2}(\eta_{2})^{-1}$ & $Q^{2}(\eta_{2})^{-1}$  & $-2$ &
$2\lambda^{2}$\\
15. Schwarzschild--(A)dS & $-\frac{1}{2}(\eta_{2})^{-1}$ & $-l\eta_{2}$ & $-2$ &
$2\lambda^{2}$\\
16. Achucarro--Ortiz & $0$ & $Q^{2}(\eta_{2})^{-1}-\frac{J}{4}(\eta_{2})^{-3}$
& $-2(\!\eta_{2}\!)^{-1}$ & $-\Lambda$\\
17. 2D Type 0A/0B & $-(\eta_{2})^{-1}$ & $\frac{BGq^{2}}{32\pi}$ & $-2(\eta_{2})^{-1}$ & $4b^{2}$\\
18. Exact String BH / Na- & $-\frac{1}{\gamma N_{\pm}(\gamma)}$ & $0$ & $-2\gamma^{-1}$ &
$4b^{2}$\\[-3pt]
ked Singularity {\scriptsize $(\eta_{2}=\gamma\pm$} & & & &\\[-3pt]
{\scriptsize $\mathrm{arcsinh}\gamma, N_{\pm}(\gamma)=1+$} & & & &\\[-3pt]
{\scriptsize $2\gamma^{-1}(\gamma^{-1}\pm\sqrt{1+\gamma^{-2}}))$} & & & &\\
19. KK Reduced CS  & $0$ & $-\frac{1}{2}(\eta_{2})^{3}$ & $2(\eta_{2})^{-1}$ & $c$\\
20. Symmetric Kink & $0$ & $\displaystyle -(\!\eta_{2}\!)^{3}\!\!\prod_{i=2}^{n}[
(\!\eta_{2}\!)^{2}\!\!-\!\phi_{i}^{2}]$
& $\displaystyle (\!\eta_{2}\!)^{-1}\!\!\prod_{i=2}^{n}[(\!\eta_{2}\!)^{2}\!\!-\!\phi_{i}^{2}]^{-1}$ &
$(\!\phi_{1}\!)^{2}$\\
21. KK Red. Conf. Flat & $0$ & $0$ & $-\frac{1}{\mathrm{sin}(\eta_{2}/2)}$ &
$-\frac{B}{4}$\\
22. Dual to Model 21 & $-\frac{\mathrm{tanh}(\eta_{2}/2)}{2}$
& $0$ & $-\frac{2}{\mathrm{sinh}(\eta_{2})}$ & $-\frac{A}{4}$\\
23. Dual to Model 19 & $-2(\eta_{2})^{-1}$ & $\frac{1}{4}(\eta_{2})^{-1}$ &
$\frac{1}{2}(\eta_{2})^{-3}$ & $M$\\
\hline
\end{tabular}
\end{center}
\end{table}

It is known that the usual dilaton models admit straightforward topological generalizations
in terms of non-standard 2d dilaton--Maxwell theories. To the best of our knowledge, the first
theories of this kind to be explored were the conformally transformed string
inspired model~\cite{cangemi2,jackiw2,jackiw3} and the extended de Sitter
gravity~\cite{cangemi2}. More recently, similar extensions have successfully been applied to more complex systems like the exact string black
hole~\cite{grumiller10} and the Kaluza--Klein reduced Chern--Simons gravity, featuring kink solutions~\cite{grumiller3} (cf. models 3, 10, 18, and 19 in Tab.~\ref{tab:selec}).

As for the EFOG's, they are a particular family
of non-standard dilaton-Maxwell models that include the aforementioned cases, besides providing
systems that have never been addressed in the literature so far. In Tab.~\ref{tab:selec}, we
work out the extensions of all dilaton models featuring in the up-to-date table of
Ref.~\cite{meyer}. The fifth column shows the identifications that must be performed in order to relate the extended theories to their unextended counterparts, for the reference case in which $\eta_{3}^{0}=1$. Note that the arbitrary integration constant $\eta_{3}^{0}$ merely rescales the free parameter $BG$, which acquires a proper physical interpretation in each model.

Since we wish to provide a physical interpretation for the central charge --- the fact that it determines a Planck length through Eq.~(\ref{eq:planck}) is an independent statement ---, the
basic criterium that we used to extend a given dilaton model was one of economy, which aimed at
identifying $BG$ with a free parameter that already existed in the original theory. Further
this should be done in such a way that we recover the original model when we integrate out the
Maxwell component. For instance, in model 8 we could have kept the original $V(\eta_{2})=
-(E/2)(\eta_{2})^{a+b}$ and added an arbitrary $W(\eta_{2})$. But if we had done this, when we
integrated the Maxwell component out we would get a dilaton model with $V(\eta_{2})=
-(E/2)(\eta_{2})^{a+b}+BG/W(\eta_{2})$, which does not recover the original model besides
introducing an extra free parameter $BG$ with an obscure physical meaning.

Since model 8 is very important and has only one free parameter, we conclude that $BG$ should
be typically interpreted as cosmological constant. In order to be consistent, for models 9--23
(which have more than one free parameter) we use the criterium of always identifying $BG$ with
the cosmological constant. For instance, in model 14 the original potential reads
$V(\eta_{2})=-\lambda^{2}+Q^{2}/\eta_{2}$. In this case we will choose to identify $BG$ with
the cosmological constant $\lambda^{2}$ and not with the charge $Q$.  By the way, this shows
that the non-standard Maxwell component that we are introducing should not be confused with
electromagnetism. This example also explains why models 1--8 have $V=0$, while models 9--23
do not.

It can be seen that $BG$ is typically identified with a global scale factor, which plays a
fundamental role in this context. For instance, in model 1
for $D=4$ we have the relation $G=EG_{N}$, where $G_{N}$ is the 4d Newton constant. But since
$E=BG$, we deduce somewhat surprisingly that $B=G_{N}^{-1}$. The substitution of the last equation in
Eq.~(\ref{eq:planck}) provides an interesting consistency check, since then the 2d Planck length $l_{P}$ is found to coincide with the
four-dimensional one (in units with $c=1$).  

A particularly interesting new class of models represents an extension
of the ab--family~\cite{grumiller1}, presenting black hole solutions (cf. model 8 in Tab.~\ref{tab:selec}). Concerning model 18, it
is important to remark that the role of the auxiliary BF-term that was employed in the original formulation of
the exact string black hole is here played by the Maxwell
component. In this case, the on-shell value of the U(1) charge $\eta_{3}^{0}$ is identified with the value of the dilaton at the
origin.

The theories described by Eq.~(\ref{eq:dilest}) are particular expressions of the more general
Poisson--Sigma models~\cite{grumiller1}, which are nonlinear generalizations of the BF
theories that have recently found application in superstring theory. In fact, performing the identifications $X^{I}\equiv (\eta_{a},\eta_{2},G\eta_{3})$ and $A_{I}\equiv (e^{a},\omega,a)$ (with $I\in\{1,2,3,4\}$), we can see that Eq.~(\ref{eq:dilest}) can be written as
\begin{equation}\label{eq:poisson}
S^{\mathrm{(PSM)}}=\frac{1}{G}\int_{M}\left(A_{I}\wedge \rmd X^{I}+\frac{1}{2}P^{IJ}(X)A_{I}\wedge A_{J}\right),
\end{equation}
in which $P^{IJ}(X)=\{X^{I},X^{J}\}$ is the Poisson tensor associated --- by the correspondence
$X^{I}\leftrightarrow (P_{a},J,I)$ --- to the nonlinear deformation of the extended Poincar\'e algebra given by
\begin{equation}\label{eq:deformpoinc}
\fl
\lbrack P_{a},J\rbrack = \sqrt{-h}\epsilon^{\verb+ +b}_{a}P_{b},\,\,  
\lbrack P_{a},P_{b}\rbrack=\epsilon_{ab}\left(\textrm{{\LARGE $\nu$}}(P^{a}P_{a},J)+\frac{BI}{W(J)}\right),\,\, \textrm{and}
\end{equation}
$\lbrack P_{a},I\rbrack  =  \lbrack J,I\rbrack=0$,
where {\LARGE $\nu$}$(\eta^{a}\eta_{a},\eta_{2})+[BG\eta_{3}]/W(\eta_{2})$ is the
potential that determines a model in Eq.~(\ref{eq:dilest}).

The generalized Jacobi identity $P^{IJ}_{\verb+ +,K}P^{KL}+\mathrm{cycl}(I,J,L)=0$ (where the subscript ``$,K$'' stands for the partial derivative $\partial/\partial X^{K}$)  can
be straightforwardly checked for Eq.~(\ref{eq:deformpoinc}). Moreover, for most physically interesting models $P^{IJ}(X)$ will be a polynomial, so that the nonlinear
algebra of Eq.~(\ref{eq:deformpoinc}) will be a finite W--algebra. Nevertheless, for each case
it should be tested whether Eq.~(\ref{eq:deformpoinc}) has the structure of a
reduced Kirillov--Poisson algebra~\cite{boer}. Note that, for $W(\eta_{2})=1$ and in the limit when {\LARGE $\nu$}$\to 0$, the
algebra of Eq.~(\ref{eq:deformpoinc}) tends to $\bar{\textrm{\i}}^{1}_{2}$
(cf. Eq.~(\ref{eq:algpoincest})) just as Eq.~(\ref{eq:poisson}) becomes the BF theory
describing the zero curvature limit of the extended de Sitter gravity~\cite{cangemi2,jackiw3}
(cf. models 3 and 10 in Tab.~\ref{tab:selec}).

\section{Concluding remarks}
\label{sec:5}
We have seen that an anomaly-free description of the two-dimensional elementary systems
introduces a noncommutative structure on the 2d Minkowski space. It is important to emphasize that this noncommutativity comes out as a non-perturbative result in our
formulation, rather than being assumed a priori. Furthermore, although we have only considered
point particles in order to establish our argument, the same feature is also expected
to show up in the case of matter fields, inasmuch as it stems from a general property of the
dynamical symmetries: the connection between the quasi-invariance of a Lagrangian
and the second cohomology group of its dynamical group.

Our analysis indicates the possibility of formulating generic noncommutative 2d dilaton gravity
theories, with usual non-twisted nonlinear gauge symmetries, in the standard commutative framework of the extended models presented in this paper (cf. Eq.~(\ref{eq:star}) and the discussion that precedes it). Note that in Sec.~\ref{sec:4} we have shown that a Moyal star product locally
describes a small noncommutativity in curved space-time for general extended dilaton models in the low
energy limit.
Although we are still researching this
issue, it is remarkable that we succesfully evade some no-go results~\cite{vassilevich}, concerning the
impossibility of constructing non-trivial deformations of 2d dilaton gravity in general. Indeed,
here we circumvent all of their assumptions, since we do not consider $\kappa$-deformations of the Poincar\'e algebra in the sense of Magueijo and Smolin (cf. Sec.~\ref{sec:3}), neither we
introduce any ad hoc Moyal star product as it was done in Ref.~\cite{vassilevich}. We remark
that Eq.~(\ref{eq:star}) provides a strong evidence for a star product which is not of the Moyal type at high energies.

On the other hand, it is noteworthy that an extra U(1) gauge sector was also required in order
to consistently define the noncommutative versions of both the Jackiw--Teitelboim and the
conformally transformed string gravity models (see~\cite{vassilevich} and the references therein). In this regard, we are particularly concerned
about the topological aspects of the aforementioned formulation of general noncommutative 2d
dilaton gravity theories, in view of the fact that the U(1) gauge field $a$, which was
introduced in our extended models, acts as a WZ potential capable of affecting the topology of
space--time (cf. Eq.~(\ref{eq:top})). 

Anyway, such a formulation is an interesting possibility, inasmuch as a Planck length is an unusual feature in two dimensions, which introduces a
natural ultraviolet cut-off that may lead to space and/or time discretization under certain
conditions still to be researched. Finally, we would like to add that the coupling of matter fields to the extended
generalized dilaton gravity theories as well as the global structure of the solutions to
the field equations of the latter are among other issues that will be addressed in a work in
progress, which will be published soon elsewhere.

\ack
This work has been institutionally supported by IFT / UNESP. The author is
grateful to J. A. Helay\"el-Neto for reading the manuscript and to D. Grumiller for useful
discussions.

\end{document}